\documentstyle[graphicx,amsmath,amsfonts]{article}
\bibliography{plain}
\title{Two-Photon Entanglement in a
Two-Mode Supersymmetric Model} \vspace{20mm}
\author{
  S. Javad Akhtarshenas
\thanks{E-mail:akhtarshenas@phys.ui.ac.ir}
\\
{\small Department of Physics, University of Isfahan, Isfahan,
Iran } }
\begin{document}
\maketitle
%\newpage

\begin{abstract}
We will study entangled two-photon states generated from a
two-mode supersymmetric model and quantify degree of entanglement
in terms of the entropy of entanglement. The influences of the
nonlinearity on the degree of entanglement is also examined, and
it is shown that amount of entanglement increase with increasing
the nonlinear coupling constant.


\end{abstract}
%\pagebreak

%\vspace{7cm}

\section{Introduction}
Perhaps, quantum entanglement is the most non-classical features
of quantum mechanics which has recently attracted much attention
although it was discovered many decades ago by Einstein, Podolsky,
 Rosen \cite{EPR} and  Schr\"{O}dinger \cite{shcro}. It plays a
central role in quantum information theory and provides potential
resources for communication and information processing
\cite{ben1,ben2,ben3}. By definition, a pure quantum state of two
or more subsystems is said to be entangled if it is not a product
of states of each components. A lot of works have been devoted to
the preparation and measurement of entangled states. Moreover the
possibility for generation of the entangled states with a fixed
photon number has been theoretically studied
\cite{duan1,duan2,coch,liu}. Duan et al described an entanglement
purification protocol which generates maximally entangled states
with fixed photon number from squeezed vacuum states or from
mixed Gaussian continuous states by the quantum non-demolition
measurement \cite{duan1,duan2}. Quantum teleportation using an
entangled source of fixed photon number has also been
theoretically investigated in \cite{coch}. Liu et al are used a
system of two coupled microcrytallites as a source with fixed
exciton number and quantified entanglement of the excitonic
states \cite{liu}. Therefore, the generation of a new entangled
source  with fixed photon number is an interesting task both from
experimental and theoretical viewpoints.

In this contribution, it is shown that a two-mode field with a
two-photon interaction can be used as a good source for
generation of entangled states with fixed photon number. We will
study entangled states generated from two degenerate bosonic
systems with fixed photon number, and we concern on quadratic
nonlinearity between modes to use Higgs algebra as the spectrum
generating algebra of the corresponding Hamiltonian
\cite{deb1,deb2}. We also restrict ourselves to the case that
total number of photons is odd. For this case, Debergh in
\cite{deb1} have shown that the corresponding Hamiltonian is
supersymmetric \cite{witten}.

A number of entanglement measures have been discussed in the
literature, such as the von Neumann reduced entropy, the relative
entropy of entanglement \cite{plenio} and the so called
entanglement of formation \cite{ben3}. In order to discus
entanglement of the states, we use von Neumann reduced entropy
which has widely been accepted as an entanglement measure for
pure bipartite states.

The organization of the paper is as follows. In section 2 we
introduce a quantum optics  model for two bosonic system with
supersymmetric feature. An analyticl solution of the Hamiltonian
is also given following the method of Ref. \cite{deb1}. In
section 3 the analytical results of section 2 are employed to
generate entangled two-photon states with fixed photon number.
Some examples are also considered in section 3. The paper is
concluded in section 4 with a brief conclusion.

\section{The two-mode supersymmetric Hamiltonian}
In this section we shall introduce and analyse  a model for
nonlinear interaction between two-mode field. Our method is based
on the analysis given in Ref. \cite{deb1}. Let us consider the
following family of Karrassiov-Klimov Hamiltonian \cite{karas}
which describes multi-photon process of scattering, i.e.
\begin{equation}\label{rs12ham1}
H=\omega_1 a_1^\dag a_1+\omega_2 a_2^\dag a_2+g(a_1^\dag)^s
a_2^r+g^\ast a_1^s(a_2^\dag)^r
\end{equation}
where $0\leq r \leq s$, $g$ is coupling constant and
$\omega_i\;(i=1,2)$ refer to angular frequencies of two-mode
field characterized by annihilation and creation operators $a_i$,
$a_i^\dag$ respectively, satisfying $[a_i,a_j^\dag]=\delta_{ij}$.
Hamiltonian (\ref{rs12ham1}) can be rewritten as
\begin{equation}\label{rs12ham2}
H=(\omega_1+\omega_2)R_0+(s\omega_1-r\omega_2)J_0+g J_{+}+g^\ast
J_{-},
\end{equation}
where \cite{deb2}
\begin{equation}
R_0\equiv\frac{1}{r+s}(ra_1^\dag a_1+sa_2^\dag a_2),
\end{equation}
and
\begin{equation} J_0\equiv\frac{1}{r+s}(a_1^\dag a_1-a_2^\dag
a_2), \qquad J_{+}\equiv(a_1^\dag)^s a_2^r,
 \qquad J_{-}\equiv
a_1^s(a_2^\dag)^r.
\end{equation}
It can be easily show that
\begin{equation}
[R_0,J_0]=[R_0,J_{\pm}]=0,
\end{equation}
and
\begin{equation}\label{comJ0Jpm}
[J_0,J_{\pm}]=\pm J_{\pm},
\end{equation}
for arbitrary values of $r$ and $s$.

It is obvious that $R_0$ is a constant of motion, and the total
photon number of the two-mode system is conserved. Moreover, the
infinite dimensional vectors
$\{|n_1,n_2\rangle=\frac{(a_1^\dag)^{n_1} (a_2^\dag)^{n_2}
}{\sqrt{n_1!n_2!}}|0,0\rangle$, $n_1,n_2=0,1,2,\cdots\}$ are
eigenvectors of $R_0$ with corresponding eigenvalues
$j=\frac{rn_1+sn_2}{r+s}$. Debergh \cite{deb1} has shown that in
order to have Higgs algebra as the spectrum generating algebra of
the Hamiltonian (\ref{rs12ham2}), we have to add to Eq.
(\ref{comJ0Jpm}) the following requirement
\begin{equation}
[J_{+},J_{-}]=2J_0+8\beta J_0^3,
\end{equation}
and have shown that \cite{deb1,deb2} this is possible only for
$r=s=2$, with parameter $\beta$ given by
\begin{equation}
\beta=-\frac{4}{4j^2+4j-2}, \qquad j=0,\frac{1}{2},1,\cdots .
\end{equation}
These values of $\beta$ lead to the relations \cite{deb3}
\begin{equation}
J_3|j,m\rangle=\frac{m}{2}|j,m\rangle,
\end{equation}
\begin{equation}
J_{\pm}|j,m\rangle=\sqrt{(j\mp m)(j\pm m +1)(j\mp m-1 )(j\pm m +2
)}|j,m\pm2\rangle,
\end{equation}
for $m=-j,-j+1\cdots,+j$. For a fixed total photon number $N$,
the vectors $|j,m\rangle$ are related to the two-mode Fock states
by
\begin{equation}
|j,m\rangle=|m_1\rangle_A|m_2\rangle_B=
\frac{(a_1^{\dag})^{j+m}(a_2^{\dag})^{j-m}}{\sqrt{(j+m)!(j-m)!}}|0\rangle_A|0\rangle_B,
\end{equation}
where $|m_1\rangle_A\otimes|m_2\rangle_B$ represent Fock state
with $m_1=j+m$ photons in mode A and $m_2=j-m$ photons in mode B.

In order to have more symmetry in Hamiltonian (\ref{rs12ham2}),
let us suppose that $\omega_1=\omega_2=\omega$, and concern on
the case that $g$ is real. In this case Hamiltonian
(\ref{rs12ham1}) reduce to
\begin{equation}\label{susyham}
\begin{array}{rl}
H & =\omega(a_1^\dag a_1+a_2^\dag a_2)+g\left((a_1^\dag)^2 a_2^2+
a_1^2(a_2^\dag)^2\right)
\\ &
=2\omega R_0+g\left(J_{+}+ J_{-}\right).
\end{array}
\end{equation}
Now, by expanding eigenvectors of (\ref{susyham}) as
$|\psi_k\rangle=\sum_{m=-j}^{m=j}C_m^{(k)}|j,m\rangle$ and using
eigenvalue equation $H|\psi_k\rangle=E_k|\psi_k\rangle$, we get
\begin{equation}\label{rec}
\begin{array}{rl}
E_kC_m^{(k)} &
=2j\omega C_m^{(k)} \\
 & +g C_{m-2}^{(k)}\sqrt{(j+m)(j+m-1)(j-m+1 )(j-m +2)}  \\
 & +g C_{m+2}^{(k)}\sqrt{(j-m)(j-m-1)(j+m+1 )(j+m +2)}.
\end{array}
\end{equation}
Moreover in order to have supersymmetric Hamiltonian, Debergh
concerned on the case that $j$ is a half-integer, which leads to
twofold degeneracy of all eigenenergies as
\begin{equation}
E_k=2\omega j+g\lambda_k, \qquad  k=1,2,\cdots,j+\frac{1}{2},
\end{equation}
where $\lambda_k$ is anyone of the $j+\frac{1}{2}$ different
solutions of \cite{deb1}
$$
[F(A_k,j,\lambda)]^2  \equiv
\left[\lambda^{j+\frac{1}{2}}-\sum_{k=1}^{j-\frac{1}{2}}A_k^2\lambda^{j-\frac{3}{2}}+
\left(\sum_{k<l,\; |k-l|\neq 2
}^{j-\frac{1}{2}}A_k^2A_l^2-A_{j-\frac{3}{2}}^2A_{j-\frac{1}{2}}^2\right)
\lambda^{j-\frac{7}{2}}\right.
$$
\begin{equation}\label{F}
\left.-\left(\sum_{k<l<p,\; |k-l|\neq 2,\;|k-p|\neq 2,\;|l-p|\neq
2 }
 ^{j-\frac{1}{2}}A_k^2A_l^2A_p^2-
\sum_{k=1}^{j-\frac{9}{2}}
 A_{k}^2A_{j-\frac{3}{2}}^2A_{j-\frac{1}{2}}^2\right)
\lambda^{j-\frac{11}{2}}\cdots\right]^2,
\end{equation}
where $A_k$ are defined by
\begin{equation}
A_k=(k(k+1)(2j-k)(2j-k+1))^{\frac{1}{2}}, \qquad k=1,2,\cdots,
j-\frac{1}{2}.
\end{equation}
Let us denote two eigenvectors of $H$ corresponding to twofold
degenerate eigenvalue $E_k$ with $|\psi_k^{(1)}\rangle$ and
$|\psi_k^{(2)}\rangle$.  Now, since Eq. (\ref{rec}) relates
coefficient $C_m^{(k)}$ to $C_{m+2}^{(k)}$ and $C_{m-2}^{(k)}$,
we can, without lose of generality, write these two orthonormal
eigenvectors belonging to eigensubspace $\varepsilon_k$ as
\begin{equation}\label{psi12jm}
\begin{array}{ll}
|\psi_k^{(1)}\rangle=\sum_{n=0}^{j-\frac{1}{2}}C_{j-2n}^{(k)}|j,j-2n\rangle,
&  C_{j-2n}^{(k)}=\frac{b_{j-2n}^{(k)}}
{\sqrt{\sum_{n=0}^{j-\frac{1}{2}}\left(b_{j-2n}^{(k)}\right)^2}},
\\
|\psi_k^{(2)}\rangle=\sum_{n=0}^{j-\frac{1}{2}}C_{j-2n-1}^{(k)}|j,j-2n-1\rangle,
& C_{j-2n-1}^{(k)}=\frac{b_{j-2n-1}^{(k)}}
{\sqrt{\sum_{n=0}^{j-\frac{1}{2}}\left(b_{j-2n-1}^{(k)}\right)^2}},
\end{array}
\end{equation}
where
\begin{equation}
\begin{array}{ll}
b_{j}^{(k)}=1, \hspace{5mm}&
b_{j-2n}^{(k)}=\frac{F(A_{2p-1},n-\frac{1}{2},\lambda_k)}{A_1 A_3
\cdots A_{2n-1}}, \quad n=1,\cdots,j-\frac{1}{2}, \\
b_{j-1}^{(k)}=1,  &
b_{j-2n-1}^{(k)}=\frac{F(A_{2p},n-\frac{1}{2},\lambda_k)}{A_2 A_4
\cdots A_{2n}}, \quad n=1,\cdots,j-\frac{1}{2},
\end{array}
\end{equation}
where function $F$ has been defined in Eq. (\ref{F}). In two-mode
Fock space representation, Eq. (\ref{psi12jm}) can be written as
\begin{equation}\label{psiAB}
\begin{array}{l}
|\psi_k^{(1)}\rangle=\sum_{n=0}^{\frac{2j-1}{2}}C_{j-2n}^{(k)}|2j-2n\rangle_A|2n\rangle_{B},
\\
|\psi_k^{(2)}\rangle=\sum_{n=0}^{\frac{2j-1}{2}}C_{j-2n}^{(k)}|2j-2n-1\rangle_A|2n+1\rangle_{B}.
\end{array}
\end{equation}
Finally, evolution operator $U(t)$ takes the following form
\begin{equation}\label{Ut}
U(t)=\sum_{k=1}^{j+\frac{1}{2}}e^{-iE_kt}
\left(|\psi_k^{(1)}\rangle\langle\psi_k^{(1)}|+|\psi_k^{(2)}\rangle\langle\psi_k^{(2)}|\right).
\end{equation}

\section{Two-photon entanglement}
In this section we will study entangled states generated by
Hamiltonian (\ref{susyham}). The entanglement measure that we are
going to use is, the so called von Neumann entropy of reduced
density matrix which has most widely been accepted as an
entanglement measure of pure state of a bipartite system.  Let
$|\psi\rangle$ be a pure state of a bipartite system with state
space $H_A\otimes H_B$. Entanglement of $|\psi\rangle$ is defined
by
\begin{equation}\label{E}
E(|\psi\rangle)=-\textmd{Tr}(\rho_A \textmd{ln}\rho_A)=
-\textmd{Tr}(\rho_B \textmd{ln}\rho_B)=\sum_n\lambda_n^2
\textmd{ln}\lambda_n^2,
\end{equation}
where $\rho_A$ is reduced density matrix of subsystem A which is
obtained by tracing out subsystem B, i.e.
$\rho_A=\textmd{Tr}_B(|\psi\rangle\langle\psi|)$, $\rho_B$ is
defined similarly, and $\lambda_n$ are square root of nonzero
eigenvalues of $\rho_A$ and $\rho_B$. They are also Schmidt
number of state $|\psi\rangle$, i.e.
\begin{equation}
|\psi\rangle=\sum_{n}\lambda_n|u_n\rangle_A|v_n\rangle_B,
\end{equation}
where $\{|u_n\rangle\}$ and $\{|v_n\rangle\}$  are orthonormal
states of two subsystems A and B, respectively. The definition is
based on the fact that although entropy of a pure state is zero,
but von Neumann entropy of each subsystem is zero only when the
state $|\psi\rangle$ is a product state.

In this paper we shall consider the case that the total number of
photons in the whole system is fixed by the initial condition
$N=2j$, and system is initially in product state
\begin{equation}\label{psi0}
|\psi(0)\rangle=|N-L\rangle_A|L\rangle_{B},
\end{equation}
where represents initially $N-L$ photons in mode A and $L$ photons
in mode B.  By taking account of Eqs. (\ref{psiAB}), (\ref{Ut}),
(\ref{psi0}), we obtain, up to an overall  phase factor
$e^{iN\omega}$, the final state of the system by
\begin{equation}\label{psit}
|\psi^{(N-L,L)}(t)\rangle=\sum_{n=0}^{\frac{N-1}{2}}
a_n|N-2n-\Delta_{L}\rangle_A|2n+\Delta_{L}\rangle_{B},
\end{equation}
where coefficients $a_n$ are defined by
\begin{equation}
a_n=\sum_{k=1}^{\frac{N+1}{2}} e^{-i\lambda_k t}\;
C_{\frac{N}{2}-L}^{(k)} C_{\frac{N}{2}-2n-\Delta_{L}}^{(k)},
\end{equation}
and $\Delta_{L}$ is difined such that it is zero (one) when $L$ is
an even (odd) integer. Obviously, Eq. (\ref{psit}) represents
final state of the system in Schmidt form and, accordingly, the
von Neumann entropy of the reduced density matrix can be obtained
easily by
\begin{equation}\label{E}
E^{(N-L,L)}(t)=-\sum_{n=0}^{j-\frac{1}{2}}
|a_n|^2\textmd{ln}|a_n|^2.
\end{equation}
Finally, it should be stress that according to Eq. (\ref{E})
maximal entangled state a of system with the total photon number
$N$ is
\begin{equation}\label{psiMAX}
|\psi_{MAX}^{(N)}\rangle=\frac{1}{\sqrt{N+1}}\sum_{n=0}^{N}
|N-n\rangle_A|n\rangle_{B},
\end{equation}
where in this case entropy of entanglement is equal to
$E_{MAX}^{(N)}=\textmd{ln}(N+1)$. On the other hand, for state
given by Eq. (\ref{psit}), maximum entropy of entanglement is
obtained when $a_n=\sqrt{\frac{2}{N+1}}$, i.e.
\begin{equation}\label{psimax}
|\psi_{MAX}^{(N-L,L)}\rangle=\sqrt{\frac{2}{N+1}}\sum_{n=0}^{\frac{N-1}{2}}
|N-2n-\Delta_{L}\rangle_A|2n+\Delta_{L}\rangle_{B},
\end{equation}
where we find $E_{MAX}^{(N-L,L)}=\textmd{ln}(\frac{N+1}{2})$.
This means that for a system with fixed photon number $N$,
maximum entanglement that can be achieved from Hamiltonian
(\ref{susyham}) is less than maximum entanglement that can be
obtained from a system that linear interaction between modes is
also considered. The difference between these two maximum is, of
course, constant and equal to $\textmd{ln}(2)$.

In the rest of this section we will consider some examples in
$N=1,3,5,9$ and discus results.
\begin{enumerate}
\item   ${\mathbf j=\frac{1}{2}}$. In this case the total number
of photons of system is $1$, and because of the twofold
degeneracy of eigenvalues, the whole state space of system
coincide with eigensubspace of the only eigenvalue. Accordingly
the final state $|\psi(t)\rangle$ differs with initial product
state only in a total phase factor, therefore, we can not have
entanglement.

\item  ${\mathbf j=\frac{3}{2}}$. In this case the state space of
system decomposes into two eigensubspces, with eigenvalues
\begin{equation}
E_{1}=3\omega+\sqrt{12}g  \qquad  E_{2}=3\omega-\sqrt{12}g.
\end{equation}
By starting with two initial states
$|\psi(0)\rangle=|3\rangle_A|0\rangle_B$ and
$|\psi(0)\rangle=|2\rangle_A|1\rangle_B$ we obtain, respectively
\begin{eqnarray}
|\psi^{(3,0)}(t)\rangle=\cos{(\sqrt{12}gt)}|3\rangle_A|0\rangle_B
-i\sin{(\sqrt{12}gt)}|1\rangle_A|2\rangle_B,
\hspace{1mm} \\
|\psi^{(2,1)}(t)\rangle=\cos{(\sqrt{12}gt)}|2\rangle_A|1\rangle_B
-i\;\sin{(\sqrt{12}gt)}|0\rangle_A|3\rangle_B.
\end{eqnarray}
By using Eq. (\ref{E}), we obtain same value for entanglement of
the above two states as
\begin{equation}\label{E30}
\begin{array}{rl}
E^{(3,0)}(t) & =E^{(2,1)}(t) \\
 & =-\cos^2{(\sqrt{12}gt)}\;\textmd{ln}(\cos^2{(\sqrt{12}gt)})
-\sin^2{(\sqrt{12}gt)}\;\textmd{ln}(\sin^2{(\sqrt{12}gt))}.
\end{array}
\end{equation}
Equation (\ref{E30}) shows that entanglement has zero value when
$t=\frac{k\pi}{2g}$ (for $k=0,1,\cdots$) and it takes maximum
value $\textmd{ln}(2)$  at times $t=\frac{(2k+1)\pi}{4g}$ (for
$k=0,1,\cdots$). This, obviously, shows that the survival time of
maximum entanglement decrease with increasing of the nonlinear
coupling constant $g$.

\item  ${\mathbf j=\frac{5}{2}}$. This case corresponds with a
system that has five photons and  the state space of
 system decomposes into three eigensubspces, with eigenvalues
\begin{equation}
E_1=5\omega   \qquad E_2=5\omega+4\sqrt{7}g  \qquad
E_3=5\omega-4\sqrt{7}g.
\end{equation}
In this case by considering the initial state as anyone of
$|\psi(0)\rangle=|5\rangle_{A}|0\rangle_{B}$,
$|\psi(0)\rangle=|4\rangle_{A}|1\rangle_{B}$ and
$|\psi(0)\rangle=|3\rangle_{A}|2\rangle_{B}$, we find,
respectively, the final state of the system as
\begin{eqnarray}\label{psit50}
\begin{array}{rl}
|\psi^{(5,0)}(t)\rangle & =
\frac{1}{14}\left(9+5\cos{(4\sqrt{7}gt)}\right)|5\rangle_{A}|0\rangle_{B}
 \\
&
-i\sqrt{\frac{5}{14}}\sin{(4\sqrt{7}gt)}|3\rangle_{A}|2\rangle_{B}
+\frac{3\sqrt{5}}{14}\left(-1+\cos{(4\sqrt{7}gt)}\right)|1\rangle_{A}|4\rangle_{B},
\end{array}
\\ \label{psit41}
\begin{array}{rl}
|\psi^{(4,1)}(t)\rangle & =
\frac{1}{14}\left(5+9\cos{(4\sqrt{7}gt)}\right)|4\rangle_{A}|1\rangle_{B}
\\
&
-i\frac{3}{\sqrt{14}}\sin{(4\sqrt{7}gt)}|2\rangle_{A}|3\rangle_{B}
+\frac{3\sqrt{5}}{14}\left(-1+\cos{(4\sqrt{7}gt)}\right)|0\rangle_{A}|5\rangle_{B},
\end{array} \hspace{1mm}
 \\  \label{psit32}
\begin{array}{rl}
|\psi^{(3,2)}(t)\rangle & =
-i\sqrt{\frac{5}{14}}\sin{(4\sqrt{7}gt)}|5\rangle_{A}|0\rangle_{B}
\\
 &  +\cos{(4\sqrt{7}gt)}|3\rangle_{A}|2\rangle_{B}
-i\frac{3}{\sqrt{14}}\sin{(4\sqrt{7}gt)}|1\rangle_{A}|4\rangle_{B}.
\end{array} \hspace{20mm}
\end{eqnarray}

\begin{figure}[t]
\centerline{\includegraphics[height=10cm]{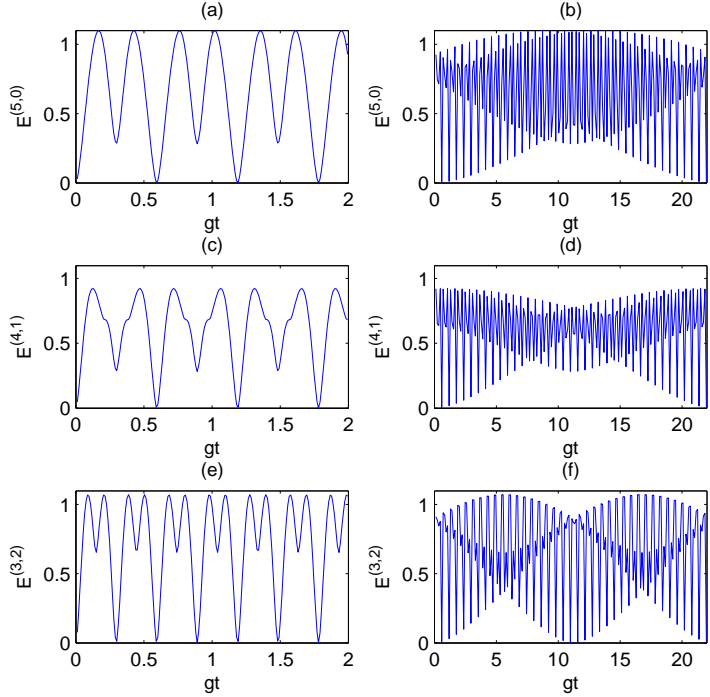}}
\caption{$E^{(5,0)}$, $E^{(4,1)}$ and $E^{(3,2)}$ are plotted as
a function of $gt$ in interval $[0,2]$ (curves $(a)$, $(c)$ and
$(e)$) and in interval $[0,22.015]$ (curves $(b)$, $(d)$ and
$(f)$).}
\end{figure}
Figure (1) demonstrates the evolution of the entropy of
entanglement as a function of $gt$ for three different initial
states with different nonlinear coupling constant $g$. The figure
is plotted such that the top horizontal line of each curve
corresponds to the maximum entanglement $\textmd{ln}(3)$. The
maximum entanglement that can be obtained by system is different
for different initial state and the system can reach,
approximately, to maximum entanglement $\textmd{ln}(3)$ only in
the case that the initial state is
$|\psi(0)\rangle=|5\rangle_{A}|0\rangle_{B}$. The Fig. (1)  also
shows that the survival time of maximum entanglement decrease
when the difference between photon numbers of two modes A and B
of the initial state is decreased. As the horizontal axis of the
curves is product of coupling constant $g$ and time $t$, it is
obvious that  by increasing the nonlinear constant $g$, survival
time of maximum entanglement decreases. Equations (\ref{psit50}),
(\ref{psit41}) and (\ref{psit32}) show that if the nonlinear
coupling constant $g$ is equal to zero, then
$|\psi(t)\rangle=|\psi(0)\rangle$, that is we can not have
entangled state.

\item  ${\mathbf j=\frac{9}{2}}$. Finally we consider as the last example the system with
nine photons and accordingly the state space of system decomposes
into five eigensubspces, with eigenvalues
\begin{equation}\begin{array}{ll}
E_1=9\omega & \\
 E_2=9\omega+\sqrt{792+24\sqrt{561}}g &
E_3=9\omega+\sqrt{792-24\sqrt{561}}g  \\
 E_4=9\omega-\sqrt{792+24\sqrt{561}}g &
E_5=9\omega-\sqrt{792-24\sqrt{561}}g.
\end{array}
\end{equation}

\begin{figure}[t]
\centerline{\includegraphics[height=8cm]{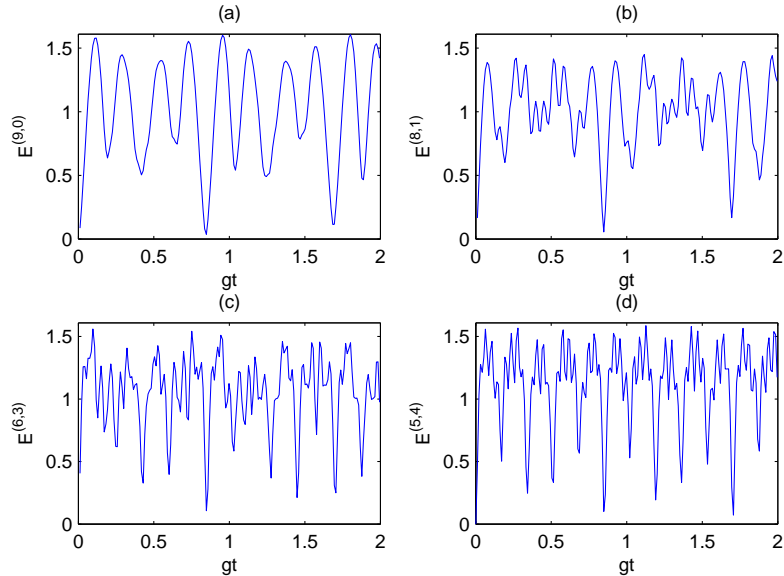}}
\caption{$E^{(9,0)}$, $E^{(8,1)}$, $E^{(6,3)}$ and $E^{(5,4)}$
are plotted as a function of $gt$ (curves $(a)$, $(b)$, $(c)$ and
$(d)$).}
\end{figure}
The evolution of  the entropy of entanglement as a function of
$gt$ for four different initial states
$|\psi(0)\rangle=|9\rangle_{A}|0\rangle_{B}$,
$|\psi(0)\rangle=|8\rangle_{A}|1\rangle_{B}$,
$|\psi(0)\rangle=|6\rangle_{A}|3\rangle_{B}$ and
$|\psi(0)\rangle=|5\rangle_{A}|4\rangle_{B}$ is demonstrated in
Fig. (2). The maximum entanglement that can be obtained by system
is different for different initial states (the top horizontal line
of each curve corresponds to the maximum entanglement
$\textmd{ln}(5)$). We find that the maximum entanglement
$\textmd{ln}(5)$ is obtained, approximately, only in the case
that there are nine photons initially in one of the modes (for
example mode A), i.e.
$|\psi(0)\rangle=|9\rangle_{A}|0\rangle_{B}$. The survival time
of the maximum entanglement decrease by increasing the nonlinear
constant $g$ and it is also decrease by decreasing the difference
between photon number of two modes A and B of the initial state.

\end{enumerate}

\section{Conclusion}
We studied entangled states generated from two-mode
supersymmetric model with fixed photon number.  We found that
only in the case that system has $N=3$ photons, the maximum
entanglement can be obtained exactly. For other systems with total
photon number greater than three, we found that the maximum
entanglement is obtained, approximately, only in the case that all
photons are initially in one of the modes, i.e.
$|\psi(0)\rangle=|N\rangle_{A}|0\rangle_{B}$ or
$|\psi(0)\rangle=|0\rangle_{A}|N\rangle_{B}$. The influences of
the nonlinearity on the degree of entanglement is also examined,
and  is shown that survival time of maximum entanglement decrease
by increasing the nonlinear coupling constant $g$. It is also
shown that the survival time of maximum entanglement decreases
when the difference between photon number of two modes A and B of
the initial state is decreased.

{\large \bf Acknowledgments}

 This work was supported by the
research department of university of Isfahan under Grant No.
831126.


\begin{thebibliography}{99}
\bibitem{EPR}{ A. Einstein, B. Podolsky and N. Rosen, }
{\em  Phys. Rev. {\bf 47}, 777 (1935).}
\bibitem{shcro}{ E. Schr\"{O}dinger, }{\em Naturwissenschaften
{\bf 23}, 807 (1935).}
\bibitem{ben1}{ C. H. Bennett, and S. J. Wiesner,}
{\em Phys. Rev. Lett. {\bf 69}, 2881 (1992).}
\bibitem{ben2}{ C. H. Bennett, G. Brassard,
C. Cr\'{e}peau, R. jozsa, A Peres and W. K. Wootters,}
{\em Phys.
Rev. Lett. {\bf 70}, 1895 (1993).}
\bibitem{ben3}{ C. H. Bennett, D. P. DiVincenzo, J. A. Smolin and W.K.
Wootters,} {\em Phys. Rev. A {\bf 54}, 3824 (1996).}
\bibitem{duan1}{ L. M. Duan, G. Giedke, J. I. Cirac and P. Zoller}{\em Phys. Rev. Lett.
{\bf 84}, 4002 (2000).}
\bibitem{duan2}{ L. M. Duan, G. Giedke, J. I. Cirac and P. Zoller}{\em Phys. Rev. A
{\bf 62}, 032304 (2000).}
\bibitem{coch}{ P. T. Cochrane, G. J. Milburn and W. J. Munro}{\em Phys. Rev. A
{\bf 62}, 062307 (2000).}
\bibitem{liu}{ Y-X Liu, S. K. \"{O}zdemir, A. Miranowicz, M. Koashi and N. Imoto }
{\em J. Phys. A: Math. Gen. {\bf 37}, 4423 (2004).}
\bibitem{deb1}{ N. Debergh, }{\em J. Phys. A: Math. Gen.
{\bf 31}, 4013 (1998).}
\bibitem{deb2}{ J. Beckers, Y. Brihaye and N. Debergh, }{\em J. Phys. A: Math. Gen.
{\bf 32}, 2791 (1999).}
\bibitem{witten}{ E. Witten, }{\em Nucl. Phys. B {\bf 188}, 513 (1981).}
\bibitem{plenio}{ M. B. Plenio and V. Vedral, }{\em Phys. Rev. A
{\bf 57}, 1619 (1998).}
\bibitem{karas}{ V. P. Karassiov and A. B. Klimov, }{\em Phys.
Lett. A {\bf 189}, 43 (1994).}
\bibitem{deb3}{ N. Debergh, }{\em J. Phys. A: Math. Gen.
{\bf 30}, 5239 (1997).}

\end{thebibliography}
\end{document}